\documentclass[pdflatex,sn-apa]{sn-jnl}

\usepackage{graphicx}%
\usepackage{multirow}%
\usepackage{amsmath,amssymb,amsfonts}%
\usepackage{amsthm}%
\usepackage{mathrsfs}%
\usepackage[title]{appendix}%
\usepackage{xcolor}%
\usepackage{textcomp}%
\usepackage{manyfoot}%
\usepackage{booktabs}%
\graphicspath{{fig/}}

\theoremstyle{thmstyleone}%
\theoremstyle{thmstyletwo}%
\theoremstyle{thmstylethree}%

\begin{document}

\title[What a Shorter Contest Really Costs]{What a Shorter Contest Really Costs:
Serial Dependence, Effective Trials, and Evidence from a Rule Change in
Professional Esports}

\author*[1]{\fnm{Petr} \sur{Parshakov}}\email{pparshakov@hse.ru}

\affil*[1]{\orgdiv{International Laboratory of Intangible-Driven Economy},
\orgname{HSE University}, \orgaddress{\city{Perm}, \country{Russian Federation}}}

\abstract{Operational research evaluates tournament formats by their efficacy, the
probability that the design identifies the stronger competitor, and computes it
almost exclusively by simulation under an assumption that is rarely tested: that
the trials a format aggregates are independent. We show that this assumption
systematically overstates what shortening a contest costs, and we measure the
correction. When trials are serially correlated, a format carries
$N/[1+(N-1)\rho]$ effective trials rather than $N$, so a twenty per cent cut costs
10.6 per cent of discriminating power under independence but only 5.2 per cent at
the correlation we estimate, and only 2.1 per cent in a sixty-trial format. The
ingredients are estimated from an implemented change in contest length: in
September 2023 professional Counter-Strike cut regulation rounds per map from
thirty to twenty-four. Across 72,576 maps we measure discrimination as the ability
slope over the residual standard deviation on the first half of a map, a contest
fixed by rule at fifteen or twelve rounds with no stopping rule. Against the
calibrated benchmark of 0.948, discrimination falls by thirty-five per cent in a
ten-month window around the switch (0.648, 95 per cent interval 0.52--0.81) and by
ten per cent over the full panel (0.901, 0.82--0.99). The correction needs one
number, recoverable from ordinary results data as a design effect.}

\keywords{Tournament design, Contest length, Efficacy, Effective sample size,
Competitive balance, Esports}

\pacs[JEL Classification]{C18, D02, L83, Z20}

\maketitle

\section{Introduction}

How much evidence a contest should collect before declaring a winner is a basic
question of tournament design. Longer contests cost players, organisers and
audiences time, while short contests let transitory shocks decide the result.
Operational research has studied this trade-off closely, and it evaluates formats
by their \emph{efficacy}: the probability that the design identifies the stronger
competitor \citep{ryvkin2008,lasek2018,sziklai2022}. Efficacy is almost always
computed by simulation, and almost always under an assumption that is convenient
rather than tested --- that the trials a format aggregates, the games, sets or
rounds, are independent.

That assumption is not innocuous. When trials are serially correlated, a format
carries fewer independent observations than its nominal length suggests, and the
marginal trial carries less than the one before it. Both effects work in the same
direction, and both are ignored by the independence benchmark. This paper measures
the dependence directly and shows what it does to the arithmetic of format design.
The answer is large and systematic: assuming independent trials \emph{overstates}
the efficacy lost by shortening a contest, by a factor of two in the setting we
study and by more in longer formats. A twenty per cent cut in trials costs 10.6
per cent of discriminating power if trials are independent, but only 5.2 per cent
once dependence is calibrated on the data, and only 3.5 per cent if the same cut
is applied to a thirty-trial contest. Section~\ref{sec:bias} sets out the general
correction; Sections~3 to 5 estimate the ingredients it needs from an implemented
change in contest length.

The empirical setting supplies what this literature usually lacks. Contest length
is rarely varied while the rest of the game stays fixed, so efficacy is theorised
and simulated far more often than it is measured.

Professional Counter-Strike provides such a variation. The game is a team shooter, and its mechanics bear directly on the analysis, so we describe them briefly. Two teams of five play a sequence of short rounds on a map. One side, the terrorists, tries to plant a bomb at one of two sites and hold it until it detonates; the other, the counter-terrorists, tries to stop the plant, defuse the bomb, or survive the clock. A round lasts roughly two minutes and ends when one side achieves its objective or eliminates the other. Teams switch roles at half time, so each plays both sides of the map. One feature is central to the analysis: equipment is bought at the start of each round with money earned in earlier ones, so a team that wins a round can afford better weapons in the next and a team that loses often cannot. Round outcomes within a map are therefore not independent draws. Series are contested over one, three or five maps taken from a seven-map pool, which the two teams narrow beforehand by alternately banning and picking.

The professional circuit is built on salaried teams and year-round tournaments, and it is large. Esports Earnings records \$163 million in prize money across more than seven thousand Counter-Strike: Global Offensive tournaments since 2012, and a further \$55 million under Counter-Strike 2 \citep{esportsearnings2026}. Valve, the publisher, sanctions two Major championships a year with a prize pool of \$1.25 million and thirty-two invited teams \citep{valve2026}; the final of the June 2026 Major drew a peak of 2.75 million concurrent viewers on Western streaming platforms alone \citep{hltv2026}. Results, statistics and a weekly world ranking of teams are published by HLTV.org, founded in 2002 and the reference archive for the discipline. Our panel records 34,453 professional matches over four years.

The competitive rules of this circuit changed in September 2023. Counter-Strike: Global Offensive, released in 2012, was replaced on 27 September 2023 by Counter-Strike 2, built on a new engine, and the competitive format moved from MR15 --- fifteen rounds a half, first team to sixteen --- to MR12, twelve rounds a half, first to thirteen, with a six-round overtime. Valve's stated reason was that changes to the economy and to weapon balance had cut the number of uncontested rounds, so matches could be settled with fewer of them. The change was global and abrupt, every round is recorded, and the internal structure of a map lets us observe a contest whose length is fixed by rule rather than terminated by a stopping rule. Because MR12 arrived with the new engine, the setting offers a sharp transition rather than an isolated experiment in length.

The change arrived with an explicit justification of exactly this kind, and with an explicit dissent. Announcing the move on 1 September 2023, Valve wrote that ``over the past decade, we've shipped updates to the economy and weapon balance to trim the fat and reduce the number of uncontested rounds in a match of CS'', and concluded that ``because of these changes, exciting competitive matches can be resolved with fewer rounds'' \citep{valve2023}. The evidence offered was about the composition of rounds rather than their number: teams reach a full buy in 73.3 per cent of non-pistol rounds after 2019 against 67.4 per cent before, and rounds in which one side cannot afford to contest have become rare. The claim is therefore not about duration but about information --- each round is said to carry enough of it that twenty-four rounds now do the work of thirty. Professional players read the same change in the opposite direction. ropz, a two-time Major winner, predicted ``even more random results'', on the grounds that pistol rounds, whose outcome can turn on the timing of a single headshot, would decide a larger share of every map \citep{hltv2023}. Both sides are making a claim about how much one round reveals about which team is better, and in the debate that followed neither measured it. That quantity is estimable, and this paper estimates it.

We measure how well a contest separates competitors by the residual-standardised ability slope, $D = \beta/\sigma_{\varepsilon}$: how far the expected outcome moves, in residual standard deviations, when the ability gap widens by one unit. The outcome is the favourite's first-half round margin, a continuous variable measured over exactly fifteen or exactly twelve rounds, with no overtime and no endogenous stopping. Ability is a pre-match Elo rating updated on round shares rather than map results, so the scale does not depend on how many rounds a map contains.

We begin with the calibration, which establishes the benchmark. Counter-Strike rounds are positively correlated through the in-game economy, and the over-dispersion of first-half scores implies an intra-class correlation of about 0.074 in both regimes. A fifteen-round half is therefore worth 7.35 independent draws, and twelve rounds of the same process 6.60. Cutting the contest by twenty per cent in rounds cuts it by ten per cent in information. The ratio of discriminations that length alone predicts is therefore 0.948, a decline of five per cent; had rounds been independent it would have been 0.894, a decline of eleven.

The observed decline exceeds this benchmark. The discrimination ratio is 0.901 over the full 2022--2025 sample, with a cluster-bootstrap interval of [0.817, 0.991]. Within a ten-month window around the transition, which compares adjacent months rather than adjacent eras, it is 0.648 [0.519, 0.811]. The full-sample estimate is statistically indistinguishable from the mechanical benchmark (p = 0.28); the window estimate is not (p < 0.001). Reweighting the new-format sample to the composition of the old moves the window estimate to 0.574, restricting to the 137 teams observed under both regimes gives 0.612, and adding team fixed effects strengthens the interaction rather than weakening it. Mechanical shortening predicts a decline of the observed sign, but it cannot account for the size of the decline observed among contemporaneous contests.

The paper makes three contributions. We measure ability revelation around an implemented change in contest length rather than simulating one. We decompose the change into an ability slope and a residual-noise component, using a statistic taken unchanged from measurement theory, which is what lets the observed decline be compared with a calibrated mechanical benchmark instead of with zero. And we separate the arithmetic a shorter contest must produce from the part it cannot, by calibrating the round process rather than assuming independent trials. We introduce no new measure: the statistic is the ordinary discrimination parameter of test theory, and Section 2.2 records its prior identities. Section 2 places these contributions against the existing work on contest length, tournament design and the measurement of ability revelation.

One limitation applies to every estimate reported below. MR12 arrived with a new engine, which changed movement, weapon behaviour, utility, the economy, the map pool and --- as our own estimates show --- the balance between the two sides. Nothing in the data separates the two changes, so the regime indicator is a joint indicator, and the results are effects of the transition rather than of round count alone. The mechanical benchmark bounds how much of the change contest shortening could produce; the excess cannot be assigned to any single accompanying mechanism.

Section 2 reviews the literature. Section 3 describes the panel and how the round format is read off each map rather than assigned by date. Section 4 sets out the discrimination statistic, the calibration of the round process and the first-half design. Section 5 reports the estimates, narrows the comparison to contests played in the same weeks, and subjects the comparison to placebo splits, alternative samples and the measures of association that the literature conventionally reports. Section 6 concludes.

\section{Related Literature}

\subsection{Contest Length, Tournament Design and the Probability that the Stronger Competitor Wins}

The closest literature treats a contest as repeated noisy evidence about relative ability. \citet{maisel1966} and \citet{gibbons1978} ask how many component games a best-of-k comparison needs before the better team wins with acceptable probability. \citet{miles1984} formalises scoring systems as symmetric sequential tests of the identity of the better player and compares their efficiency against the amount of play required. The comparative static is theirs: when component performances are informative but noisy, additional observations raise selection accuracy, by an amount that depends on the scoring and stopping rules.

Applied work makes this proposition concrete and adds a qualification. \citet{urban2013} shows that longer playoff series need not favour the better team once the home-away sequence changes with series length. In tennis, \citet{harris2016} simulates a fall of 3.5 to 5 per cent in the stronger player's win probability under no-ad and match-tiebreak scoring, and \citet{kovalchik2018} report upset frequencies of 20 per cent under Fast4 against 13 per cent in best-of-three and 10 per cent in best-of-five; \citet{simmonds2018} reach the same conclusion. \citet{dominicy2013} show that table tennis's move from 21-point to 11-point sets makes victories against the odds more likely, and \citet{percy2009} compares badminton's old and new scoring systems. These are the closest conceptual precedents for MR15 against MR12. They share a limitation: each translates assumed point- or rally-winning probabilities into counterfactual match outcomes, rather than estimating how an implemented change altered the empirical relation between ability and performance.

A second literature evaluates formats by how reliably they select or rank competitors under noise. \citet{ryvkin2008} call the probability that the ex ante best player wins a tournament its predictive power. \citet{lasek2018} compare league formats against latent team strengths, and \citet{sziklai2022} rank knockout, group, multi-stage and Swiss designs under limited comparison budgets. Selecting a winner and ranking a field are distinct objectives, and the value of an extra contest turns on how comparisons are allocated, not on their number alone.

Related work treats a single match as signal extraction. \citet{skinner2009} ask how often the better soccer team wins and how misleading an observed result can be. \citet{wolfson2018} measure how much a game reveals about team quality and how well it predicts later games. \citet{lopez2018} separate latent strength from game-level noise in a Bayesian state-space model and simulate how postseason parity varies with series length. \citet{holzmeister2025} put a magnitude on the same decomposition at season length: across seven European football leagues and 7,240 matches they attribute roughly 40 per cent of the variation in performance relative to expectation to skill and 60 per cent to luck, which over a 38-match season leaves the identity of the champion sensitive to chance with probability 0.34 and the relegation places with probability 0.76. Their finding is the complement of ours in an instructive way: they hold the format fixed and show how little of the final table survives the noise even after a long season, whereas we hold the sport fixed and vary the length. Read together, the two say that the amount of play a format prescribes is a weak instrument for the reliability it delivers. We adopt this logic --- evaluate a format by the information its outcomes carry about ability --- but apply it within a single contest rather than across a tournament, and to observed rather than simulated variation. A separate strand supplies the design rather than the question. \citet{apesteguia2010} and \citet{krumer2020} study the penalty shoot-out precisely because its length and environment are fixed by regulation while the match around it is not, and \citet{malueg2010} test contest theory on the internal structure of a best-of-three rather than on who won it. The motivation for looking inside a match rather than at its result is documented on both sides: play is not independent of the running score \citep{klaassen2001,lagopenas2014}, and the match stops as soon as one side reaches the target. Section 4.2 applies that logic to the half of a Counter-Strike map.

\subsection{Measurement of Ability Revelation and the Remaining Gap}

The statistic is not ours. The measurement literature contains it under several names, and we adopt it in the form that literature established. Classical test theory defines reliability as the share of observed-score variance attributable to stable true-score differences \citep{lord1968}. Signal-detection theory standardises the separation between two outcome distributions by their within-distribution noise \citep{stanislaw1999,massof1987,decarlo2023}. In regression, signal variance relative to residual variance is Cohen's $f^{2}$ \citep{cohen1988}. For the outcome $M = \alpha + \beta S + \varepsilon$, our $D = \beta/\sigma_{\varepsilon}$ is the performance response to a one-unit rise in ability measured in residual standard deviations. When $S$ is standardised in a bivariate model $D^{2}$ equals $R^{2}/(1-R^{2})$, reliability is $D^{2}/(1+D^{2})$, and for a contrast between two fixed ability levels $\beta \Delta S / \sigma_{\varepsilon}$ is the discriminability $d'$. The first of these identities is the signal-to-noise ratio of classical test theory, $\rho/(1-\rho)$, whose equivalence with the information function of item response theory is established by \citet{nicewander1993}; $D$ with a standardised $S$ is therefore also Cohen's $f$, and $D$ as we report it is $f$ expressed per 100 rating points rather than per standard deviation of ability. We keep D rather than R-squared because the latter also moves with the cross-sectional dispersion of ability, which differs across our two regimes. The statistic itself is standard; its application to a contest-format transition is not.

The identity that matters most for the validity of what follows is with item response theory. Under a two-parameter normal-ogive model written in its latent-response form, the discrimination parameter of an item is exactly the slope of the latent-ability response divided by the residual standard deviation, $a = \lambda/\sqrt{1-\lambda^{2}}$ for a standardised latent variable, and the information the item contributes to the ability estimate is proportional to its square \citep{ferguson1949,jordan2019,pavlech2025}. \citet{ferrando2012} develops the same family of discrimination indices for the linear factor-analytic model, which is the case we are in: our outcome is continuous and our specification is linear, so the ratio $\beta/\sigma_{\varepsilon}$ is the discrimination index of that model rather than an approximation to a binary-response one. Read a contest as an item that measures the latent ability of the competitors and D is that discrimination parameter, while the effective trial count of Section 4.2 is the information the contest carries. The correspondence is not an analogy constructed for this paper. Scale-invariance in the outcome, sensitivity to the units of the latent variable, and accumulation with the number of conditionally independent trials are properties established in that literature; we inherit them rather than assert them, and the interpretation of a cross-regime ratio of discriminations follows from them directly. The same ratio is what the tournament-design literature varies when it evaluates formats: the predictive power of \citet{ryvkin2008} is a monotone transformation of the gap in ability divided by the standard deviation of performance noise, so a format that raises $D$ is a format that more often selects the better competitor.

The correction we apply for serial dependence is equally standard. The ratio of the observed variance of a half score to its binomial counterpart is the design effect of survey sampling, and the nominal number of rounds divided by it is the effective sample size \citep{kish1965}. The same quantity appears in sports economics as the square of the Noll-Scully ratio of the actual to the idealised standard deviation of win percentages, whose dependence on the number of games played is well documented \citep{owen2010}. \citet{doria2021} make that dependence the object of study and construct measures of competitive balance that are invariant to season length, on the grounds that a measure moving mechanically with the amount of play cannot be compared across leagues that play different amounts. Our problem is the within-league form of theirs: because the treatment here is a change in the amount of play, a statistic that responds mechanically to length must be given a length-only benchmark before any residual movement can be read as substantive. Section 5.2 constructs that benchmark from the data rather than assuming independence, which is what the simulation literature on scoring systems typically assumes.

Two clarifications keep the statistic where it belongs. First, discrimination in our sense is not the discriminatory power of contest theory. The parameter r of a Tullock contest success function, and the noise parameters that generalise it, govern how sensitively a modelled win probability responds to effort and are properties of an assumed functional form \citep{amegashie2006,jia2008}; D is an estimated property of realised outcomes, and nothing in our design identifies r. Second, the empirical literature has measured objects of the same family without using this name. \citet{lopez2018} and \citet{wolfson2018} separate team strength from game-level noise; \citet{heuer2020} scores performance indicators by how much of the underlying team strength they recover net of statistical noise; \citet{aoki2017} and \citet{jerdee2024} decompose outcomes into skill and luck, the latter with an explicit depth-of-competition parameter; \citet{bennaim2013} work directly with the upset probability. \citet{miles1984} and its generalisation by \citet{pollard2015} evaluate scoring systems by the amount of play required to reach a given probability of identifying the better competitor, which is the same trade-off expressed as a cost rather than as a slope. We use D because it is estimable regime by regime from contests that were actually played and because its comparative static in length is sharp; we do not claim it improves on any of these.

Evidence from implemented length changes remains scarce. \citet{dimattia2026} exploit a 2017 reform that cut round-robin matches per team in beach volleyball from three to two, and find no detectable change in favourite-win probability or in the alignment of final standings with initial strength. Theirs is the closest observed structural-change precedent, but the treatment varies the number of matches in a tournament stage rather than the number of repeated trials inside a single contest, and the outcome remains a win indicator.

Esports research on tournament design largely imports methods built for traditional sport, and work on Counter-Strike concentrates on prediction, player valuation and in-game decisions. We found no scholarly analysis of the MR15-to-MR12 transition. The gap this paper fills is therefore not the proposition that shorter contests favour weaker competitors, which is well established. It is the measurement of ability revelation around an implemented transition, the separation of the ability-response and residual-noise components, and the comparison of the observed change with a mechanical benchmark calibrated on the effective number of rounds. Every element of the measurement apparatus is borrowed. What is new is that it is applied to a change in contest length that actually occurred, and that it is given a benchmark against which the estimate can fail.

\section{Data}

\subsection{Source and Sample Construction}

The panel is assembled from HLTV.org, the reference results archive for professional Counter-Strike. It covers 1 January 2022 to 31 December 2025. The pipeline requests only URL families permitted by the site's robots.txt: the paginated results archive, individual match pages, the weekly world-ranking archive and event pages. The date-filtered results view is disallowed, so the study window is located by binary search over the offset parameter of the unfiltered results list and then traversed page by page. Every page is stored in a compressed raw cache and every request is logged, so all downstream results can be re-derived without re-contacting the source. The crawl retrieved 34,453 match pages with no failures and no skipped observations.

The window opens on 1 January 2022 rather than earlier for a substantive reason. Competitive play through 2020 and much of 2021 ran under pandemic restrictions: matches were played online, international travel was suspended for long stretches, and the circuit was segmented into regional brackets. Venue is a first-order determinant of play in this discipline --- 18.1 per cent of our maps are at a LAN event and the indicator enters every map-level specification --- and regional segmentation breaks the network of common opponents on which any rating recursion depends. 2022 is the first full year after the circuit returned to its normal calendar. The cost of that choice is that the rating recursion begins with the sample rather than before it, so ratings are least informative in the opening months of the panel. Section 5.3 documents the consequence and shows that removing those months moves the estimate away from one rather than towards it, so the full-sample figure is the conservative one.

The unit of observation is a played map. A best-of-three series ending 2-0 contributes two observations, not three; the unplayed slot stays in the database but never enters the estimation sample. The raw panel contains 72,576 played maps and 17,820 unplayed slots across 34,453 matches, 2,547 teams and 2,451 events. Teams are identified by persistent site identifiers rather than names, so renamings and sponsor changes do not fragment a team's history.

Three exclusions define the estimation sample. We drop 1,446 walkover placeholders: a forfeit is recorded as a nominal 1-0 result on a map named ``Default'' with no round detail, and such records are not contests. We drop observations for which the pre-match world ranking of either team is unavailable, because the favourite cannot be defined; this affects 13,131 maps, overwhelmingly matches between teams outside the published ranking. We drop the small number of maps whose round format cannot be determined. The estimation sample contains 58,294 maps from 26,958 matches, 1,094 teams and 1,934 events. Specifications using the half decomposition also require the half-by-half score and the starting side to be recoverable and both teams to have at least twenty-five prior maps in the rating recursion, leaving 39,546 maps from 19,362 matches.

\subsection{Construction of the Key Variables}

The regime variable is the round format of the map: MR15, fifteen rounds per half and first to sixteen, or MR12, twelve rounds per half and first to thirteen. Assigning the format by calendar date relative to the engine release on 27 September 2023 would introduce substantial measurement error. Several leagues finished their previous-generation seasons after that date, and some final scores are observationally identical across regimes: a map ending 16-14 is either an MR15 regulation result or an MR12 map decided in the first overtime.

The format is instead inferred from the scoreline of the map itself. The source reports the half-by-half score of every map in the form (10:5; 6:8), and a half is always played to completion, so the rounds contested in the first half identify the regime directly. Where the half string is missing we infer the format from unambiguous final scores, then from the observed format of other maps in the same event, and only as a last resort from the calendar; each observation records which rule applied. In the estimation window 97.96 per cent of maps are classified from the half structure, 1.92 per cent from the event majority, 0.04 per cent from the final score and 0.08 per cent from the calendar. A consistency filter discards half evidence that cannot generate the map's own final score; five maps are affected.

The favourite is the team with the better official world ranking in force before the match. We collected the weekly ranking archive in full for 29 November 2021 to 29 December 2025: 214 consecutive snapshots and 53,211 team-week observations, with roughly 235 to 265 ranked teams per snapshot. Snapshots are located by following each page's own previous-week link rather than by generating Mondays, which picks up two irregular publication gaps a calendar-based iterator would miss.

Matches are linked to rankings by an as-of join with a strict no-look-ahead rule: the applicable snapshot is the latest published on a strictly earlier calendar day, and each team is looked up in that snapshot only, so a team absent from it counts as unranked rather than carrying forward a stale rank. The realised lag runs from one to eight days, median four. Both teams are ranked for 81.9 per cent of maps.

One institutional caveat is in order. The HLTV ranking is the consensus reference in the discipline, but it is not the ranking Valve uses: Major invitations and seeding run off the Valve Regional Standings. We use the HLTV ranking because it is published weekly, archived back to 2012 and covers far more teams than the invitation lists, and because it defines the favourite rather than allocating anything. Nothing in the design requires the ranking to be institutionally binding; Section 4.3 introduces a continuous rating estimated from the data itself, and the main results use that.

\subsection{Descriptive Statistics and Data Availability}

\begin{table}[!ht]
\caption{Descriptive statistics for the estimation sample.}
\label{tab:desc}
\centering
\footnotesize
\begin{tabular}{lrrrr}
\toprule
 & \multicolumn{2}{c}{Transition window} & \multicolumn{2}{c}{Full sample} \\
 & \multicolumn{2}{c}{Jun 2023 -- Mar 2024} & \multicolumn{2}{c}{2022--2025} \\
\cmidrule(lr){2-3}\cmidrule(lr){4-5}
 & MR15 & MR12 & MR15 & MR12 \\
\midrule
\multicolumn{5}{l}{\emph{Coverage}}\\
Maps                                  & 2,579 & 4,202 & 12,229 & 27,317 \\
Matches                               & 1,261 & 2,110 & 6,178  & 13,184 \\
Distinct teams                        & 177   & 192   & 284    & 400 \\
Distinct events                       & 126   & 210   & 592    & 953 \\
\addlinespace
\multicolumn{5}{l}{\emph{The contest: first half of a map}}\\
Rounds per half (by rule)             & 15    & 12    & 15     & 12 \\
Favourite's rounds won, mean          & 7.96  & 6.39  & 7.97   & 6.42 \\
\hspace{1em} standard deviation       & 2.71  & 2.32  & 2.80   & 2.35 \\
Favourite's round share               & 0.530 & 0.532 & 0.531  & 0.535 \\
Favourite's margin, mean              & 0.91  & 0.77  & 0.94   & 0.85 \\
\hspace{1em} standard deviation       & 5.42  & 4.64  & 5.59   & 4.70 \\
Halves ending level                   & 0.000 & 0.139 & 0.000  & 0.135 \\
\addlinespace
\multicolumn{5}{l}{\emph{Ability and matchups}}\\
Rating gap, mean (points)             & 21.2  & 20.1  & 16.4   & 23.1 \\
\hspace{1em} median                   & 19.7  & 19.3  & 15.4   & 21.6 \\
\hspace{1em} 90th percentile          & 58.7  & 60.3  & 51.2   & 66.3 \\
World-rank gap, median                & 24    & 24    & 22     & 24 \\
Favourite's world rank, median        & 45    & 42    & 42     & 49 \\
\addlinespace
\multicolumn{5}{l}{\emph{Setting}}\\
Favourite starts on CT                & 0.510 & 0.507 & 0.505  & 0.503 \\
Played at a LAN event                 & 0.195 & 0.184 & 0.206  & 0.211 \\
Best-of-one                           & 0.037 & 0.110 & 0.080  & 0.077 \\
Best-of-three                         & 0.947 & 0.875 & 0.904  & 0.902 \\
Map won by the lower-ranked team       & 0.417 & 0.408 & 0.413  & 0.401 \\
\bottomrule
\end{tabular}

\smallskip
\begin{minipage}{\textwidth}
\footnotesize
\emph{Notes.} One observation is one played map for which the half decomposition
and the starting side are recoverable and both teams have at least twenty-five
prior maps. Walkovers, forfeits and maps with an incomplete pre-match ranking are
excluded. The rating gap is the pre-match round-share Elo difference, favourite
minus underdog, in rating points. The favourite is the better-ranked team at the
time of the match. A fifteen-round half cannot end level; a twelve-round half can,
and does in about one case in seven.
\end{minipage}
\end{table}

Table \ref{tab:desc} describes the estimation sample, window first. Two features
matter for what follows. The composition of matchups is close to identical across
regimes --- median rating gaps of 19.7 and 19.3 points inside the window, the same
median world-rank gap, near-identical shares of LAN play and of best-of-three
series --- so the comparison is not between different kinds of match. And the
unconditional rate at which the lower-ranked team wins the map differs by under one
percentage point between regimes. Neither fact settles anything: the regimes are separated in calendar time, and the map result is in any case a truncated and endogenously stopped statistic, which is why Section 4.2 works with the first half instead.

Appendix~\ref{secA1} documents the crawl, the rules by which the round format is read off each map, the exclusions and the construction of every variable; a replication package contains the pipeline, the parsers, the database schema, the estimation code and the compressed page cache.

\section{Empirical Framework}

\subsection{A Measure of Performance Discrimination and the Prediction of Contest Length}

Let the outcome of a contest between two competitors be

\begin{equation}
M = \alpha + \beta S + \varepsilon ,
\end{equation}

where S is the difference in their ability and $\varepsilon$ collects everything the contest itself contributes. We define the discrimination of the contest as

\begin{equation}
D \;=\; \frac{\beta}{\sigma_{\varepsilon}} .
\end{equation}

The two ingredients have plain readings. The slope $\beta$ is the \emph{signal}: how far the expected outcome moves when one competitor is better than the other. The residual standard deviation $\sigma_{\varepsilon}$ is the \emph{noise}: how far realised outcomes scatter around that expectation when the ability gap is held fixed. In the estimates below the signal is 3.16 rounds per hundred rating points and the noise 5.42 rounds, so the contest is noisy: a single map is a weak signal of which team is better, which is why the question of how many rounds to play has force at all. The intercept $\alpha$ is a level term and plays no part in $D$; Section 4.2 explains why it is not zero. D answers the question a contest designer faces: by how many noise standard deviations does the expected outcome move when the ability gap widens by one unit. Two properties of D are relevant for what follows. D is invariant to the scale of the performance outcome, so first-half margins measured over fifteen and twelve rounds can be compared directly. It is not invariant to the units of ability, so the comparison requires S to stay on a common pre-match rating scale across regimes --- which is why the rating recursion in Section 4.3 updates on round shares rather than map results. Unlike an R-squared or a correlation, D does not rise merely because the sample happens to contain a more dispersed set of ability gaps. That distinction is not hypothetical in this application: Section 5.5 shows the dispersion of the ability gap rising by a quarter across the transition, which is enough to reverse the sign of every standardised measure.

What makes $D$ testable rather than merely descriptive is that it has a sharp
comparative static in contest length. If a contest aggregates $n$ conditionally
independent trials whose per-trial informativeness is unchanged, the signal
accumulates in proportion to $n$ while the noise accumulates in proportion to
$\sqrt{n}$, so
\begin{equation}
D(n) \;\propto\; \sqrt{n} .
\label{eq:sqrtn}
\end{equation}
The prediction for a change in length is therefore a ratio of square roots of
\emph{effective} trial counts, and the effective count is estimable from the data.
Section 4.2 estimates it. (The identities that make $D$ a standard object rather
than one of our own are set out in Section 2.2 and not repeated here.)

Let the stronger side win each round with probability p greater than one half. A race to sixteen over thirty regulation rounds resolves in favour of the stronger side more often than a race to thirteen over twenty-four, for every p, so the discrimination of the shorter contest is lower. The sign is unambiguous. The magnitude is not: it depends on how much independent information a single round carries.

One question the design invites is why the half should be affected at all. It is
because the half is not a fixed object across the two regimes: it is fifteen rounds
under MR15 and twelve under MR12, so the treatment applies directly to the contest
we measure. The magnitude then follows from \eqref{eq:sqrtn}. If rounds were
independent, twelve of them would carry the information of twelve and fifteen the
information of fifteen, and $D$ would fall by the factor $\sqrt{12/15} = 0.894$ ---
a decline of 10.6 per cent, not the 20 per cent that counting rounds suggests.
Serial dependence moves the prediction closer to one rather than away from it,
because the rounds that were removed were partly redundant; Section 4.2 measures how
much, and Section 5.2 carries the calculation out. The point to carry forward is
that the null of interest is a benchmark strictly between 0.894 and one, and never
zero: nothing about the competitors has to change for a shorter contest to
discriminate less.

\subsection{The Round-Generating Process and the First-Half Specification}

Rounds within a map are not independent. Equipment bought in one round persists into the next, so wins arrive in streaks. We model this as a beta-binomial process: the per-round win probability of a given map is drawn once from a Beta distribution centred on the team's true value, and rounds are conditionally independent given that draw. The intra-class correlation is identified from the over-dispersion of first-half scores relative to the binomial benchmark, computed within forty fine strata of pre-match strength so that heterogeneity between matchups is not mistaken for correlation within a map:

\begin{equation}
\varphi \;=\; \frac{\operatorname{Var}(X)}{N\,\bar p\,(1-\bar p)},
\qquad
\rho \;=\; \frac{\varphi - 1}{N - 1},
\qquad
N_{\mathrm{eff}} \;=\; \frac{N}{\varphi},
\end{equation}

where $X$ is the favourite's first-half round count, $\bar p$ its mean share and $N$ the number of rounds in the half. $N_{\mathrm{eff}}$ is the number of independent draws a half is worth, and the point prediction for the ratio of discriminations across regimes is $\sqrt{N_{\mathrm{eff},12}/N_{\mathrm{eff},15}}$.

The design exploits a feature of the rules. The first half of a map is a contest of exactly fifteen rounds under MR15 and exactly twelve under MR12. It has no stopping rule, no overtime and a continuous outcome, the round margin. It carries the same 12/15 length ratio as regulation play, 24 rounds against 30, without the truncation and the discrete threshold that make the map result an inefficient statistic. For each regime we estimate

\begin{equation}
M_i \;=\; \alpha + \beta S_i + \gamma\,\mathit{CT}_i + \delta_{m(i)\times \mathrm{side}} + \varepsilon_i ,
\label{eq:half}
\end{equation}

where $M_i$ is the favourite's first-half round margin, $S_i$ the pre-match rating difference, $\mathit{CT}_i$ an indicator for the favourite starting on the counter-terrorist side, and $\delta_{m(i)\times\mathrm{side}}$ a set of map fixed effects interacted with the side indicator, which absorbs the differing side balance of individual maps. Discrimination is $\beta$ divided by the residual standard deviation, and the statistic of interest is the cross-regime ratio.

This follows a literature that isolates a sub-contest whose length and conditions are fixed by the rules, rather than analysing the match that contains it. \citet{apesteguia2010} and \citet{krumer2020} use the penalty shoot-out for that reason: the number of attempts is set by regulation, the environment is common to both sides, and the order is drawn by lot, none of which is true of the match around it. \citet{malueg2010} test contest theory on the internal structure of a best-of-three rather than on its outcome, and \citet{magnus1999} analyse a single named set of a tennis match as an object with its own properties. The reason for not using the whole match is documented from two directions: behaviour responds to the running score, so play recorded after the score moves is contaminated \citep{klaassen2001,lagopenas2014}, and the match itself ends the moment one side reaches the target, which censors the outcome and makes the amount of play a function of the state. The first half of a map is the counterpart of the shoot-out in this setting --- fixed at fifteen rounds under MR15 and twelve under MR12, unable to end early, played by both teams under a side assignment we observe --- with one addition: the outcome is a continuous margin rather than a binary result, which carries more information about strength \citep{kovalchik2020,lopez2018}. We follow the logic of these papers, not their question: none of them measures what happens when the length of the contest is changed.

A second, pooled specification interacts ability with the regime indicator:

\begin{equation}
M_i \;=\; \alpha + \beta S_i + \theta\,\mathit{MR12}_i + \delta\,(S_i \times \mathit{MR12}_i)
+ \gamma\,\mathit{CT}_i + \lambda_{\mathrm{map}\times\mathrm{side}} + \mu_t + \varepsilon_i ,
\label{eq:pooled}
\end{equation}

with $\mu_t$ a set of calendar fixed effects at the week or month level. The parameter of interest is $\delta$, the change in the ability slope; $\theta$ is a nuisance term. The pooled form admits calendar and team fixed effects but forces the residual variance to be common across regimes, so the two specifications complement each other: the ratio of discriminations lets the noise term differ, the interaction lets us control time and teams.

\subsection{Measurement of Ability}

The published ranking is coarse: integer-valued, weekly, and covering only the leading two hundred and fifty or so teams. We build a continuous rating from the data. Ratings follow an Elo recursion with logistic scale 400 and all teams entering at 1500, updated after every map by

\begin{equation}
r_a \;\leftarrow\; r_a + K\!\left(\frac{s_a}{s_a + s_b}
- \frac{1}{1 + 10^{(r_b - r_a)/400}}\right) ,
\end{equation}

where $s_a$ and $s_b$ are rounds won. Two features of this recursion matter. The update uses the round share rather than the map result, so it does not depend on how many rounds a map contains and ratings estimated under MR12 sit on the same scale as ratings estimated under MR15. And every map is scored with the ratings both teams carried into it, so the regressor is strictly pre-determined.

The implementation is deliberately simple. Maps are processed in chronological order by date, then match identifier, then map number. There is no seasonal decay, no roster-change adjustment and no special handling of inactive teams. New teams enter at the common prior and are excluded from estimation until they have twenty-five prior maps, which is also required of their opponent. Walkovers carry no round information and do not update ratings. We fixed the step size K = 24 before estimating any reported specification, but chose it with reference to in-sample predictive fit rather than a pre-registered rule, so Table 5 reports the main statistic across K in \{12, 18, 24, 32, 48\}.

\subsection{Identification}

The round format did not change in isolation. MR12 arrived with a new engine, which altered movement, weapon behaviour, utility, the economy, the map pool and, as Section 5.5 documents, the balance between the two sides. Nothing reachable under the source's access rules carries an independent per-match marker of the engine, and the handful of leagues that continued under MR15 after the release is the only variation separating the two. The regime indicator is therefore a joint indicator for the format change and the engine change, and the estimates are effects of the transition, not of round count net of everything else.

This has an implication for how the results should be interpreted. If the observed change in discrimination matched the magnitude implied by the reduction in rounds alone, that would be consistent with a pure length mechanism but would not prove one, since an engine effect of zero is not testable here. If the observed change were substantially larger, the excess would have to come from the accompanying changes. Section 5.3 finds the second pattern.

What the design does deliver is comparability along two dimensions that would otherwise be fatal. The first-half outcome fixes contest length by rule rather than by the state of play, so the comparison is not contaminated by the endogenous stopping rule of the map. And the transition-window specifications hold calendar time fixed, which removes the confound of a professional scene that changed considerably between 2022 and 2025. Neither device removes the engine change.

\subsection{Inference}

Maps of the same series are not independent, so we cluster standard errors at the match level throughout. The discrimination ratio is a non-linear function of two fits estimated on disjoint samples, so we obtain its sampling distribution by a cluster bootstrap over matches, drawn separately within each regime, with 1,000 replications for the headline estimates, 600 for the regime-specific levels of $D$ plotted in Figure \ref{fig:levels}, 400 for the remaining rows of Table \ref{tab:calendar} and 300 for the alternative measures of Table \ref{tab:measures}; the smaller counts are a concession to the cost of refitting eight statistics per draw, and doubling them moves no reported interval endpoint by more than 0.005. Reported p-values are two-sided bootstrap probabilities of a ratio at least as far from the stated null as the estimate. The calibrated benchmark of Section 5.2 is itself estimated, but with a bootstrap interval of [0.946, 0.949] it is treated as known when the observed ratios are tested against it.

\section{Empirical Results}

\subsection{Discrimination Before and After the Change}

Table 2 reports the first-half specification, estimated separately in each regime.
Both ingredients of $D$ move as the mechanism implies. The ability slope falls from
3.16 to 2.41 rounds per hundred rating points: a given difference in strength now
shows up as a smaller lead. The residual standard deviation falls from 5.42 to 4.59
rounds: outcomes scatter less as well, because there are fewer rounds to scatter
over. Neither movement is informative on its own. What matters is which of them
fell faster.

\begin{table}[!ht]
\caption{Ability and the first-half round margin, by round format.}
\label{tab:main}
\centering
\footnotesize
\begin{tabular}{lcccc}
\toprule
 & \multicolumn{2}{c}{Transition window} & \multicolumn{2}{c}{Full sample} \\
\cmidrule(lr){2-3}\cmidrule(lr){4-5}
 & (1) MR15 & (2) MR12 & (3) MR15 & (4) MR12 \\
\midrule
Rating gap ($\beta$, rounds per 100 points)
  & 3.772 & 2.109 & 3.164 & 2.414 \\
  & (0.367) & (0.222) & (0.183) & (0.084) \\
\addlinespace
Favourite starts on CT ($\gamma$)
  & 0.487 & 0.197 & 1.484 & 0.072 \\
  & (0.325) & (0.211) & (0.171) & (0.093) \\
\addlinespace
Map $\times$ starting-side fixed effects & yes & yes & yes & yes \\
\addlinespace
Residual standard deviation ($\sigma_\varepsilon$, rounds)
  & 5.225 & 4.513 & 5.418 & 4.585 \\
$R^2$ & 0.077 & 0.055 & 0.062 & 0.049 \\
Maps & 2,579 & 4,202 & 12,229 & 27,317 \\
\addlinespace
\textbf{Discrimination} $D=\beta/\sigma_\varepsilon$
  & \textbf{0.722} & \textbf{0.467} & \textbf{0.584} & \textbf{0.526} \\
95\% cluster-bootstrap interval
  & [0.588, 0.869] & [0.372, 0.566] & [0.520, 0.650] & [0.486, 0.560] \\
\addlinespace
$D(\text{MR12})/D(\text{MR15})$
  & \multicolumn{2}{c}{\textbf{0.648} \; [0.519, 0.811]}
  & \multicolumn{2}{c}{\textbf{0.901} \; [0.817, 0.991]} \\
\bottomrule
\end{tabular}

\smallskip
\begin{minipage}{\textwidth}
\footnotesize
\emph{Notes.} The dependent variable is the favourite's first-half round margin,
in rounds. Standard errors in parentheses are clustered at the match level.
Intervals for $D$ and for the ratio come from a cluster bootstrap over matches
that recomputes $\beta$, $\sigma_\varepsilon$ and their ratio in each replication;
the interval on the ratio is bootstrapped on the ratio itself and is not implied
by the two separate intervals.
\end{minipage}
\end{table}

\begin{figure}[htbp]
\centering
\includegraphics[width=\textwidth]{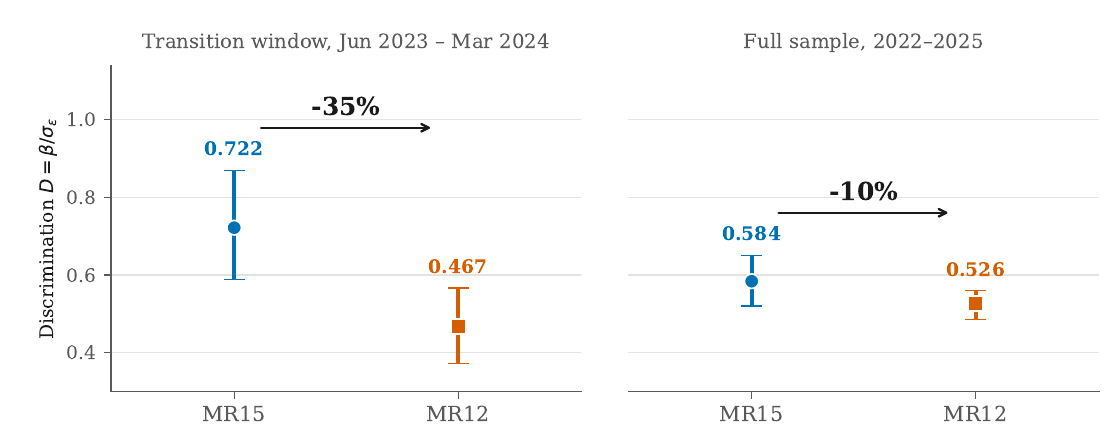}
\caption{Discrimination before and after the change, window first. Points are $D = \beta/\sigma_{\varepsilon}$ from the specification of Table \ref{tab:main}, estimated separately by regime; bars are 95 per cent cluster-bootstrap intervals over matches, 600 replications. The two panels have separate vertical scales in common but are otherwise independent comparisons: levels are not comparable across panels, only the within-panel ratios are.}
\label{fig:levels}
\end{figure}

Figure \ref{fig:levels} and Table \ref{tab:main} put the answer in one picture and
one table, and we read them window first.

Inside the transition window --- June 2023 to March 2024, the ten months on either
side of the switch --- discrimination falls from 0.722 to 0.467. That is a
reduction of a third, and the two cluster-bootstrap intervals barely touch. This is
the comparison the paper is built on, and it is worth being exact about what it
does and does not deliver.

What it delivers is calendar \emph{proximity}, not simultaneity. Inside the window
the two formats remain largely sequential: MR15 maps run from June to October 2023
and MR12 maps from October 2023 to March 2024, because leagues switched as their
seasons ended rather than mid-season. Of the forty-three weeks in the window only
three contain maps of both formats, and only three of the 333 events do. The window
therefore removes the multi-year gap that separates the regimes in the full sample
--- it compares adjacent months rather than adjacent eras --- but it does not hold
the calendar fixed, and any shock common to the autumn of 2023 is not differenced
out by it.

The genuinely contemporaneous comparison is the last row of Table
\ref{tab:calendar}: the 139 MR15 and 344 MR12 maps played in those three overlapping
weeks, with week fixed effects. Its point estimate is 0.492, further from one than
the window estimate, but its interval is [0.213, 1.103] and covers no change. It is
the cleanest design in the paper and the weakest evidence in it, and we report it as
a directional check rather than as a result.

Over the full 2022--2025 sample the same statistic falls from 0.584 to 0.526, a
reduction of about ten per cent, and the intervals overlap. The full sample is
larger --- 12,229 and 27,317 maps against 2,579 and 4,202 --- and therefore more
precise, but it is also the weaker design: it compares a mostly 2022--2023 MR15
group with a mostly 2024--2025 MR12 group, and much besides the format changed in
between. We report it throughout as the estimate from the larger and more
distant comparison rather than as the estimate of interest. It is not a bound in
either direction: nothing in the design makes the full-sample number a floor under
the true effect, and if the professional scene changed in ways that raised
discrimination over 2024--2025 it could equally overstate it. What the two numbers
jointly show is a pattern --- every restriction that moves the comparison closer in
calendar time moves the ratio further from one, which is the sequence in Table
\ref{tab:calendar} --- and a pattern of that shape is hard to generate from an era
effect, though it does not rule one out.

Expressed as a ratio, discrimination under MR12 is 0.648 of its MR15 value inside
the window, with an interval of [0.519, 0.811], and 0.901 [0.817, 0.991] on the
full sample. Those two numbers are the object of the rest of the paper. They cannot
be read, however, against a null of one: a shorter contest must discriminate less
even if nothing else changed at all. The next subsection works out by how much.

\subsection{What the Change Should Have Cost}

The question has an exact answer, and reaching it takes two steps: first what a
shorter contest costs when its rounds are independent, then how much of that cost
survives the fact that they are not. Both steps are standard; only the second
requires data.

\textbf{Step 1: if rounds were independent.} Write the margin as an advantage plus luck.
The advantage points the same way in every round and therefore accumulates in
proportion to the number of rounds: at a per-round win probability of 0.55, fifteen
rounds give an expected margin of $15 \times (0.55-0.45) = 1.50$ rounds and twelve
give 1.20. Luck points either way and therefore partly cancels: in the sum of $N$
independent deviations the cross-terms average to zero, so it is the squared
deviations that add and the standard deviation grows as $\sqrt{N}$ rather than as
$N$. Over fifteen rounds the standard deviation of the margin is
$2\sqrt{15 \times 0.55 \times 0.45} = 3.85$ rounds; over twelve it is 3.45. The
signal-to-noise ratio is therefore $1.50/3.85 = 0.389$ against $1.20/3.45 = 0.348$,
and their ratio is

\begin{equation}
\frac{D(12)}{D(15)} \;=\; \frac{0.1 \times 12 / \sqrt{12}}{0.1 \times 15/\sqrt{15}}
\;=\; \frac{\sqrt{12}}{\sqrt{15}} \;=\; \sqrt{\tfrac{12}{15}} \;=\; 0.894 .
\label{eq:independent}
\end{equation}

The per-round advantage cancels between numerator and denominator, which is what
makes this a usable benchmark: it depends on how many rounds were removed and on
nothing about the teams.

The illustration just given compares a fixed pair of teams, and it is worth being
explicit that the object it constructs is the one we estimate. Our $D$ is
$\beta/\sigma_{\varepsilon}$, in which $\beta$ is a derivative with respect to the
rating gap rather than the expected margin of a particular pair. Let
$\delta = \partial p/\partial S$ denote the rise in the favourite's per-round win
probability produced by a hundred rating points. Under $N$ independent rounds,
$\mathbb{E}[M \mid S] = N(2p-1)$ and $\operatorname{Var}(M \mid S) = 4Np(1-p)$, so
\begin{equation}
\beta = \frac{\partial \mathbb{E}[M \mid S]}{\partial S} = 2N\delta,
\qquad
\sigma_{\varepsilon} = 2\sqrt{Np(1-p)},
\qquad
D(N) = \frac{\delta}{\sqrt{p(1-p)}}\,\sqrt{N} .
\label{eq:Dscaling}
\end{equation}
The first factor collects everything about the competitors and is common to the
two regimes, so it drops out of any ratio taken at a common $p$, leaving
$\sqrt{N_{12}/N_{15}}$ exactly as in \eqref{eq:independent}. Reading $D$ instead
as the signal-to-noise ratio of one fixed pair,
$\mathbb{E}[M]/\operatorname{SD}(M) = (p-\tfrac12)\sqrt{N}/\sqrt{p(1-p)}$, changes
the level but not the scaling: in both readings the numerator is proportional to
$N$ and the denominator to $\sqrt{N}$. The benchmark is therefore invariant to
which of the two quantities the reader has in mind, and take a stronger favourite
and every quantity in the calculation changes except the ratio. Neither the square-root law nor its use in evaluating
formats is new. It is the content of the Spearman-Brown prophecy formula
\citep{spearman1910,brown1910,lord1968}, under which lengthening a test raises
its reliability and its discrimination with the square root of its length, and it
is the mechanism behind the predictive power of \citet{ryvkin2008}. What the
present setting adds is a length change that actually happened, and a round
process whose departure from independence can be measured rather than assumed.

One consequence of \eqref{eq:Dscaling} organises what follows. Since
$\sigma_{\varepsilon}$ is the standard deviation of the margin \emph{conditional
on} the rating gap, the variance inflation that replaces $N$ by an effective count
must be measured conditionally as well; this is why the calibration of Section
5.2 is computed within strata of the rating gap rather than on the pooled
distribution of half scores.

\textbf{Step 2: but rounds are not independent.} Equipment bought in one round persists
into the next, so wins arrive in streaks, and fifteen rounds are worth fewer than
fifteen independent observations. The dependence is not measured round by round; it
is measured through the trace it leaves in the spread of half scores. Two features of the favourite's
round count $W$ must be kept apart here. Its \emph{mean} is the advantage: at the
observed mean share of $\bar p = 0.531$, $\mathbb{E}[W] = 7.97$ against the
underdog's 7.03, a margin of about one round in fifteen. Its \emph{standard
deviation} is the spread, and that is the quantity the calibration uses. If rounds
were independent draws at $\bar p$, then $W \sim \text{Binomial}(15,\bar p)$ and
that spread would be
$\sqrt{15 \times 0.531 \times 0.469} = 1.93$ rounds --- how far from 7.97 a typical
half lands, not an advantage of 1.93 rounds --- and scores such as 13--2 would be
rare. Figure 2 shows what the data contain instead.

\begin{figure}[htbp]
\centering
\includegraphics[width=\textwidth]{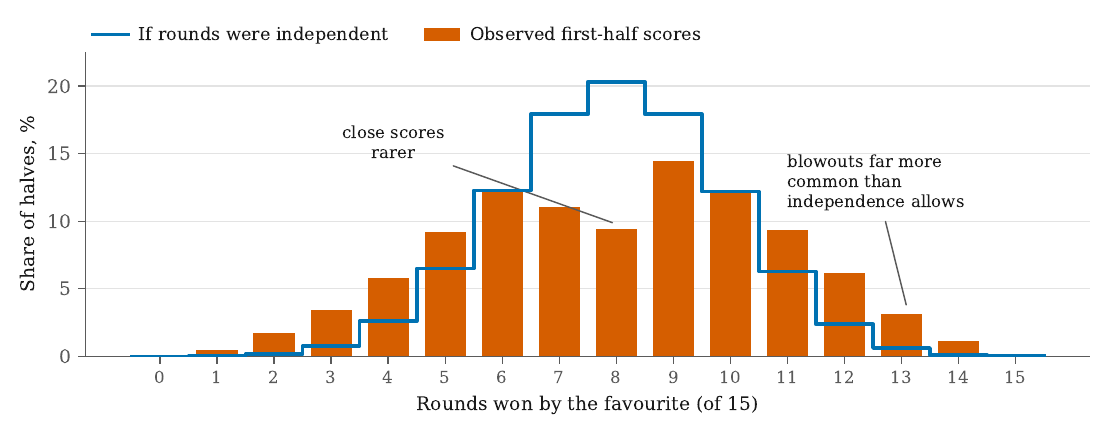}
\caption{Dependence between rounds, read off the spread of half scores. Bars are the observed distribution of the favourite's first-half round count in the 12,229 MR15 maps of the estimation sample; the step function is the binomial distribution implied by independent rounds at the same mean share. Close scores are half as common as independence allows and blowouts four times as common.}
\label{fig:clumping}
\end{figure}

Scores of 8--7 and 7--8 account for 20.5 per cent of halves against the 38.2 per
cent independence implies, and scores of twelve or more (or three or fewer) for
16.2 per cent against 4.0. The realised standard deviation is 2.77 rounds against
the 1.93 of independence, larger by a factor of 1.43, or by

\begin{equation}
\varphi \;=\; \frac{\operatorname{Var}(X)}{N\,\bar p\,(1-\bar p)} \;=\; 1.43^{2} \;=\; 2.05
\label{eq:phi}
\end{equation}

These figures are on the scale of the round count. Because a margin doubles a
count, $M = 2W - N$, the same two quantities on the margin scale of Section 4.1
are 3.86 rounds under independence against 5.54 realised, the latter being the
object the regression estimates as $\sigma_{\varepsilon} = 5.42$; the small
difference is the controls. Since $\varphi$ is a ratio of variances, it is
identical on either scale.

in squared terms. Squared terms are the right metric because that is what
accumulates across rounds, as equation~(\ref{eq:independent}) already used: each
independent round adds exactly one unit to the squared spread, so the squared
spread divides evenly by the number of rounds and the linear spread does not.

Two details make $\varphi$ an estimate of dependence rather than of something else.
First, the spread is computed within forty fine strata of the pre-match rating
difference: pairs of unequal strength produce spread of their own, and without
strata that would be counted as clumping. The correction is small here --- 2.09
without strata against 2.05 with them --- because the observation is already
oriented as favourite against underdog, but it has to be shown rather than assumed.
Second, simulating the round process with the implied intra-class correlation
reproduces what the data contain: half scores drawn with $\rho = 0.075$ have a
standard deviation of 2.76 rounds and produce blowouts in 16.0 per cent of halves,
against 2.77 and 16.2 in the sample, while independent rounds give 1.93 and 4.0.

Dividing the nominal count by $\varphi$ gives the effective number of independent
rounds: $15/2.05 = 7.32$ under MR15 and $12/1.80 = 6.65$ under MR12. In the
vocabulary of survey sampling $\varphi$ is Kish's (1965) design effect and $N/\varphi$
the effective sample size; $\sqrt{\varphi}$ is the Noll-Scully ratio of the
competitive-balance literature \citep{owen2010,doria2021}.

\begin{table}[!ht]
\caption{Calibration of the round-generating process, by round format.}
\label{tab:calib}
\centering
\footnotesize
\begin{tabular}{lcc}
\toprule
 & MR15 & MR12 \\
\midrule
Rounds in the half, $N$                                   & 15    & 12 \\
Favourite's mean round share, $\bar p$                    & 0.531 & 0.535 \\
\addlinespace
Spread of the favourite's round count if rounds were independent, & & \\
\hspace{1em} $\sqrt{N\bar p(1-\bar p)}$                    & 1.93  & 1.73 \\
Spread actually observed (within strata of the rating gap) & 2.77  & 2.32 \\
\addlinespace
Over-dispersion $\varphi$                                  & 2.05  & 1.80 \\
Implied intra-class correlation $\rho = (\varphi-1)/(N-1)$ & 0.075 & 0.073 \\
\textbf{Effective rounds} $N_{\text{eff}} = N/\varphi$     & \textbf{7.32} & \textbf{6.65} \\
\addlinespace
Maps                                                       & 12,229 & 27,317 \\
\bottomrule
\end{tabular}

\smallskip
\begin{minipage}{\textwidth}
\footnotesize
\emph{Notes.} Over-dispersion is computed within forty strata of the pre-match
rating gap, because $\sigma_\varepsilon$ is a spread conditional on ability and
the correction must be conditional as well; the pooled figures are 2.09 and 1.85.
Because a margin doubles a count, $M = 2W - N$, the same two spreads on the margin
scale of Section 4.1 are 3.86 and 5.54 rounds for MR15, the latter being the object
estimated as $\sigma_\varepsilon = 5.42$ in Table \ref{tab:main}; $\varphi$ is a
ratio of variances and is identical on either scale. The nominal counts differ by
20 per cent; the effective counts differ by 9.2 per cent.
\end{minipage}
\end{table}

The two effective counts differ less than the two nominal counts do, and the reason
is the crux of the exercise. Dependence makes each additional round add less than
the one before: with $\varphi = 1 + (N-1)\rho$ and $\rho$ around 0.074, effective
rounds rise from 4.4 at $N=6$ to 7.3 at $N=15$ and only to 9.5 at $N=30$. The three
rounds removed from each half were the cheapest ones.

A length-only counterfactual must hold the round process fixed and vary only $N$.
We therefore fix $\rho$ at its MR15 value, $\hat\rho = 0.0743$, and ask what
twelve rounds of \emph{that same} process would carry:
$\varphi(12) = 1 + 11\hat\rho = 1.817$ and $N_{\text{eff}}(12) = 6.60$, against
$N_{\text{eff}}(15) = 7.35$. Twenty per cent fewer rounds are then ten per cent less
information, and because discrimination grows with the square root of information
rather than in proportion to it, the predicted fall is

\begin{equation}
\frac{D(\mathrm{MR}12)}{D(\mathrm{MR}15)} \;=\; \sqrt{\frac{N_{\text{eff}}(12)}{N_{\text{eff}}(15)}}
\;=\; \sqrt{\frac{6.60}{7.35}} \;=\; 0.948 .
\label{eq:benchmark}
\end{equation}

Three remarks on this number, since it is the paper's main construction.

First, the three candidate predictions should be stated together rather than
contrasted with a straw man. Twenty per cent of the rounds were removed. If rounds
were independent, $D$ would fall by 10.6 per cent, to 0.894. Under the dependence
actually measured it falls by 5.2 per cent, to 0.948. Dependence moves the
prediction \emph{towards} no change, because redundant rounds were the ones lost.

Second, the benchmark is not sensitive to $\rho$ in a way that threatens the
comparison, and its own sampling uncertainty is small. Setting $\rho$ to 0, 0.04,
0.10 and 0.20 gives benchmarks of 0.894, 0.931, 0.956 and 0.975: over the entire
plausible range the prediction stays above 0.89, so no value of $\rho$ produces
anything near the window estimate of 0.648. A cluster bootstrap over matches of the
MR15 calibration puts the benchmark at 0.948 with a 95 per cent interval of
[0.946, 0.949], two orders of magnitude tighter than the interval on the observed
ratio. Tests in Section 5.3 therefore treat the benchmark as known; treating it as
estimated would widen the comparison by a negligible amount.

Third, allowing each regime its own $\rho$ instead --- 0.0743 under MR15 and 0.0726
under MR12 --- gives 0.953 rather than 0.948. We prefer the fixed-$\rho$ version
because a counterfactual in which only the length changes should not import the new
regime's dependence structure, and we note that the two versions differ by half a
percentage point, far inside the uncertainty of anything they are compared with.
That the two estimates of $\rho$ are so close is itself reassuring: had the round
process changed materially, the benchmark would be measuring the engine rather than
the format.

\subsection{The Observed Change Against the Benchmarks}

Figure 3 places the two estimates on the same scale as the three benchmarks.

\begin{figure}[htbp]
\centering
\includegraphics[width=\textwidth]{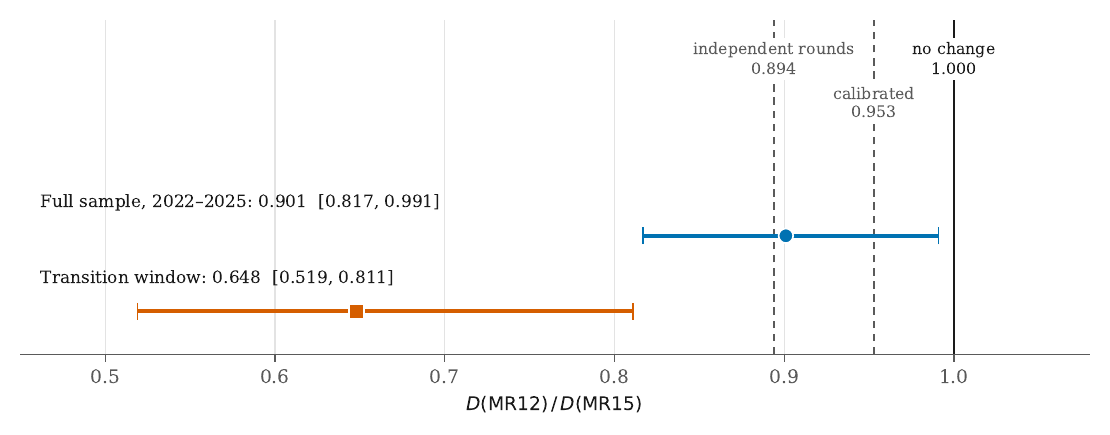}
\caption{The observed decline against what shortening alone implies. Points are the ratio of regime-specific discriminations with 95 per cent cluster-bootstrap intervals over matches; vertical lines are the prediction under independent rounds, the prediction under the round dependence actually observed, and no change at all.}
\label{fig:benchmarks}
\end{figure}

On the full sample the estimate of 0.901 sits below the calibrated benchmark but
its interval contains it: against a null of 0.948 the bootstrap p-value is 0.31. On
this comparison alone the mechanical channel cannot be rejected. Inside the
transition window the estimate of 0.648 lies below both benchmarks, and the bootstrap p-value against 0.948 is below 0.001. The
difference between the two is calendar distance, and Table \ref{tab:calendar} varies it systematically.

\begin{table}[!ht]
\caption{Discrimination ratios under alternative restrictions in calendar time.}
\label{tab:calendar}
\centering
\footnotesize
\begin{tabular}{lrrcc}
\toprule
Sample & MR15 maps & MR12 maps & $D$ ratio & Pooled interaction \\
\midrule
\textbf{Transition window, Jun 2023 -- Mar 2024}
  & 2,579 & 4,202 & \textbf{0.648} & $-1.599$ \\
  &       &       & [0.519, 0.811] & (0.429) \\
\hspace{1em} adding favourite-team fixed effects
  &       &       &                & $-1.813$ \\
  &       &       &                & (0.443) \\
\addlinespace
Both teams played under both formats
  & 2,114 & 3,072 & 0.612 & \\
  &       &       & [0.462, 0.819] & \\
Sep -- Dec 2023 only
  & 925   & 2,305 & 0.695 & \\
Weeks containing both formats, week fixed effects
  & 139   & 344   & 0.492 & $-1.652$ \\
  &       &       & [0.213, 1.103] & (0.432) \\
\addlinespace
Full sample, 2022--2025
  & 12,229 & 27,317 & 0.901 & $-0.762$ \\
  &        &        & [0.817, 0.991] & (0.201) \\
\bottomrule
\end{tabular}

\smallskip
\begin{minipage}{\textwidth}
\footnotesize
\emph{Notes.} The $D$ ratio is the ratio of regime-specific discriminations from
the specification of Table \ref{tab:main}, with 95 per cent cluster-bootstrap
intervals over matches in brackets (1,000 replications for the first and last rows,
400 elsewhere). The pooled interaction is the coefficient on the rating gap
interacted with the MR12 indicator, in rounds per hundred rating points, with
match-clustered standard errors in parentheses. All pooled specifications include
the starting-side indicator and map-by-side fixed effects. Tightening the
comparison moves the ratio further from one, not towards it.
\end{minipage}
\end{table}

The estimate does not attenuate as the comparison tightens; it moves further from
one. In the preferred window the ratio is 0.648 [0.519, 0.811] and the pooled
interaction is negative and significant. Adding favourite-team fixed effects moves
that interaction from $-1.599$ to $-1.813$, so the contrast is not produced by a
change in which teams are being compared. Of the 232 teams appearing in the window,
137 are observed under both formats; restricting to the 5,186 maps in which both
participants belong to that set gives 0.612 [0.462, 0.819]. The final row uses only
the three weeks in which both regimes were played, with week fixed effects: 139
MR15 maps, a point estimate of 0.492 and an interval of [0.213, 1.103] that
includes no change. It is the cleanest comparison in principle and the weakest in
practice, and we report it as a directional check rather than as evidence in its
own right.

\subsection{What the Change Means for Who Wins}

Ratios of discriminations are the right object for testing and the wrong object for
a rulebook. This subsection translates them into units an organiser can use. It is
an illustration computed from the fitted model rather than a further estimate, and
two things about it need stating before the numbers.

\textbf{What the quantity is.} Under the specification of Section 4.1 the margin is
normal with mean $\alpha + \beta S$ and standard deviation $\sigma_{\varepsilon}$,
so
\begin{equation}
\Pr(M > 0) \;=\; \Phi\!\left(\frac{\alpha + \beta S}{\sigma_{\varepsilon}}\right)
\;=\; \Phi(D S),
\label{eq:winprob}
\end{equation}
where the second equality sets the level term to the value at which equally rated
teams split the half evenly. A continuous approximation, however, cannot be read as
the probability of \emph{winning} a twelve-round half, because such a half can end
six--six and 13.9 per cent of them do; a fifteen-round half cannot be drawn. The
quantity $\Phi(DS)$ is instead the favourite's \emph{share of the decision} --- a
win counted as one and a draw as one half --- which is the only version of the
statistic that is comparable across a format that admits draws and one that does
not.

That reading is not an assertion. Evaluating the calibrated beta-binomial of
Section 4.2 at the same rating gaps gives the favourite a share of 53.6, 57.2, 60.7
and 63.2 per cent at gaps of 20, 40, 60 and 74 rating points, against 53.7, 57.4,
61.0 and 63.5 from $\Phi(DS)$; the discrete and the continuous versions agree to
within half a point across the range. Table \ref{tab:winprob} reports both, so that
the outright-win and draw probabilities are visible alongside the share. One
caveat: at the window's matchups the beta-binomial puts 15.8 per cent of MR12
halves level against 13.9 per cent observed, so it somewhat overstates the draw
mass and correspondingly understates outright wins.

\textbf{Where the numbers are evaluated.} The rating gaps in Table
\ref{tab:winprob} are quantiles of the gap distribution actually observed in the
window, not round numbers. The median gap is 19.5 rating points, the 75th
percentile 39.7, the 90th 59.9 and the 95th 73.6; fewer than one map in a hundred
exceeds 100 points and the largest gap in the window is 153. Evaluating the
expression at 200 or 300 points, as an earlier draft of this section did, is
extrapolation beyond the support of the data and we do not report it.

\begin{figure}[!ht]
\centering
\includegraphics[width=\textwidth]{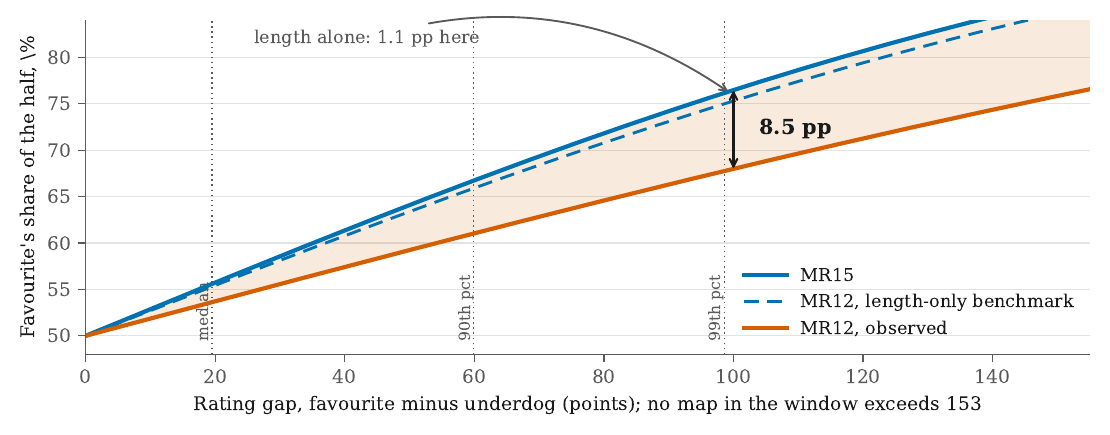}
\caption{The favourite's share of the half, by rating gap, over the range the
window contains. Curves use the window estimates: MR15, the counterfactual under a
length-only change, and MR12 as observed. Dotted lines mark the median, 90th and
99th percentiles of the observed gap distribution; the largest gap in the window is
153 points. The shaded area is what the transition cost the better team.}
\label{fig:winprob}
\end{figure}

Read that way, Table \ref{tab:winprob} and Figure \ref{fig:winprob} say three
things.

First, what shortening alone can explain is small in these units. Across the
observed range of gaps the length-only benchmark costs the favourite at most 1.3
points of share, and at the median gap about 0.3. An organiser choosing between
twelve and fifteen rounds, and right in believing nothing else would move, would be
trading roughly one decision in three hundred at a typical matchup for a fifth of
the playing time. On those terms the trade is defensible.

Second, the change that occurred is an order of magnitude larger. At the 90th
percentile of the gap the favourite's share falls from 66.8 to 61.0 per cent, and at
the median from 55.7 to 53.7. The largest gap anywhere in the window, 153 points,
gives 10.3 points of share. These are not dramatic numbers, and it is better to state
them plainly than to reach for a more impressive gap that the data do not contain:
what the transition cost the better team, at the matchups professional
Counter-Strike actually produces, is on the order of two to six points of decision
share.

Third, the effect is hump-shaped in the rating gap and the observed range sits on
its rising limb. Between evenly matched teams the outcome is near a coin flip under
any format; the cost of a shorter contest grows with the gap over the entire range
present in the data. Whether it would turn down again at the very large gaps where
the favourite wins under any format is a property of the functional form, not
something the window can show.

\begin{table}[!ht]
\caption{What the change cost the better team, at rating gaps the window actually contains.}
\label{tab:winprob}
\centering
\footnotesize
\begin{tabular}{lcccc@{\hskip 1.6em}ccc}
\toprule
& \multicolumn{4}{c}{Favourite's share of the half (per cent)}
& \multicolumn{3}{c}{Discrete model} \\
\cmidrule(lr){2-5}\cmidrule(lr){6-8}
Rating gap & MR15 & Length-only & MR12 & Cost & MR15 & MR12 & MR12 \\
(points)   &      & benchmark   & observed & (pp) & wins & wins & drawn \\
\midrule
\phantom{0}20 \; (median)      & 55.7 & 55.4 & 53.7 & 2.0 & 55.5 & 45.6 & 16.0 \\
\phantom{0}40 \; (75th pct)    & 61.4 & 60.8 & 57.4 & 4.0 & 60.9 & 49.3 & 15.8 \\
\phantom{0}60 \; (90th pct)    & 66.8 & 65.9 & 61.0 & 5.7 & 66.1 & 53.0 & 15.4 \\
\phantom{0}74 \; (95th pct)    & 70.3 & 69.4 & 63.5 & 6.8 & 69.7 & 55.6 & 15.1 \\
\bottomrule
\end{tabular}

\smallskip
\begin{minipage}{\textwidth}
\footnotesize
\emph{Notes.} The first four columns are $\Phi(DS)$ from equation
\eqref{eq:winprob}, evaluated at the window estimates of $D$, with the level
normalised so that equally rated teams split the half evenly. This is the
favourite's share of the decision: a win counted as one, a drawn half as one half.
The benchmark column applies the length-only ratio of 0.948 to the MR15 estimate.
``Cost'' is the MR15 share minus the MR12 share, in percentage points. The last
three columns come from the calibrated beta-binomial of Section 4.2 at the same
gaps and are stated as outright probabilities; a fifteen-round half cannot be
drawn. The beta-binomial puts slightly more mass on drawn halves than the data do
(15.8 against 13.9 per cent unconditionally), so it understates outright MR12 wins
by two to four points. Rating gaps are quantiles of the distribution observed in
the window; the largest gap in the window is 153 points, where the cost reaches
10.3.
\end{minipage}
\end{table}

A related question --- how much additional play would restore the lost
discrimination --- can be answered within the same model, with a warning attached.
Because $D$ grows with the square root of the effective number of rounds, restoring
a ratio $r$ requires multiplying effective rounds by $r^{-2}$. For the length-only
benchmark the factor is 1.11, which takes the effective count back to its MR15 value
and, inverting $N_{\text{eff}} = N/[1+(N-1)\rho]$, requires about fifteen nominal
rounds. That the calculation returns the format we started from is a check on the
calibration rather than a new result. For the full-sample ratio of 0.901 the factor
is 1.23 and the required half length is about nineteen rounds; for the window ratio
of 0.648 it is 2.38, and the required effective count of 15.7 exceeds $1/\rho
\approx 13.5$.

That last comparison should be read as a statement about the fitted model and not
about Counter-Strike. The ceiling $1/\rho$ follows from the constant pairwise
correlation of the beta-binomial, which was calibrated on halves of twelve and
fifteen rounds; nothing in the data speaks to whether a fiftieth round would be
correlated with the first in the same way, and a dependence structure that decays
with distance between rounds would have no such ceiling. Within the calibrated
model, lengthening the half cannot recover the window's decline; outside it, the
honest statement is only that the required lengthening is large and that recovering
discrimination across contests --- more maps in a series --- is a cheaper instrument
than lengthening one.

\subsection{Robustness}

\begin{table}[!ht]
\caption{Sensitivity of the discrimination ratio to specification and sample.}
\label{tab:robust}
\centering
\footnotesize
\begin{tabular}{llc}
\toprule
 & Variant & $D$ ratio \\
\midrule
\emph{Benchmark}    & Transition window (Table \ref{tab:calendar}) & 0.648 \\
                    & Full sample                                  & 0.901 \\
\addlinespace
\emph{Rating recursion} & $K = 12$  & 0.869 \\
                    & $K = 18$  & 0.890 \\
                    & $K = 24$ (baseline) & 0.901 \\
                    & $K = 32$  & 0.907 \\
                    & $K = 48$  & 0.904 \\
\addlinespace
\emph{Composition}  & Window reweighted to the MR15 matchup distribution & 0.574 \\
                    & Teams observed under both formats                  & 0.612 \\
\addlinespace
\emph{Sample start} & From Jan 2022 (baseline) & 0.901 \\
                    & From Jul 2022            & 0.843 \\
                    & From Jan 2023            & 0.854 \\
                    & From Apr 2023            & 0.842 \\
\addlinespace
\emph{Burn-in}      & $\geq 25$ prior maps (baseline) & 0.901 \\
                    & $\geq 50$ prior maps  & 0.899 \\
                    & $\geq 100$ prior maps & 0.909 \\
\addlinespace
\emph{Placebo splits} & MR12, split at Oct 2024 & 0.981 \; [0.850, 1.139] \\
                    & MR12, split at Jan 2025 & 0.945 \; [0.816, 1.101] \\
                    & MR15, split at Jan 2023 & 0.926 \; [0.749, 1.151] \\
                    & MR15, split at Mar 2023 & 0.945 \; [0.789, 1.179] \\
\bottomrule
\end{tabular}

\smallskip
\begin{minipage}{\textwidth}
\footnotesize
\emph{Notes.} Reweighting matches the MR12 sample to the MR15 joint distribution of
rank-gap quintile, favourite-rank quintile, venue and series length, keeping the 76
cells common to both. The start-date rows re-estimate the full-sample ratio on maps
played from the stated date onwards; the burn-in rows raise the number of prior maps
each team must have. The placebo rows split a single regime at a date on which no
rule changed and report the ratio of the later half to the earlier one; all four
intervals cover one, so the procedure does not manufacture declines where no rule
changed.
\end{minipage}
\end{table}

\textbf{The narrow window.} The window comparison rests on 2,579 MR15 and 4,202
MR12 maps, and its interval is correspondingly wide. It is nonetheless the
specification we prefer, because the alternative is a comparison across eras. The
rows below all ask whether something other than the format produces the contrast.

\textbf{The rating recursion.} The step size of the Elo update was fixed at
$K = 24$ before any reported estimate, but chosen with reference to in-sample fit.
Across a fourfold range --- $K$ = 12, 18, 24, 32, 48 --- the full-sample ratio is
0.869, 0.890, 0.901, 0.907 and 0.904. Nothing here turns on that choice.

\textbf{Composition.} Reweighting the MR12 sample of the window to the MR15 joint
distribution of matchup asymmetry, favourite strength, venue and series length
moves the ratio from 0.648 to 0.574; restricting to teams observed under both
formats gives 0.612. Conditioning harder on composition moves the estimate away
from one, which is the opposite of what a composition story predicts.

\textbf{Pre-trends.} Discrimination by quarter runs 0.62, 0.68, 0.62, 0.51 and 0.73
within MR15 from the third quarter of 2022 to the third of 2023, and between 0.44
and 0.63 across nine quarters of MR12. Interacting the rating difference with a
linear month index inside each regime gives a drift of $+0.066$ rounds per hundred
points per month under MR15 (standard error 0.036) and $+0.008$ under MR12 (0.011).
The MR15 drift is upward, which is what a warming-up rating produces, and it
depresses the early MR15 slope: it pushes the full-sample ratio towards one rather
than away from it.

\textbf{The start of the sample.} The panel opens in January 2022 for the reason
given in Section 3.1, and the two lowest quarters in it are the first two of 2022,
at 0.34 and 0.29, when the rating recursion has only just started. Moving the start
forward to July 2022, January 2023 and April 2023 moves the full-sample ratio from
0.901 to 0.843, 0.854 and 0.842; raising the minimum number of prior maps from
twenty-five to fifty and to one hundred leaves it at 0.899 and 0.909. Every
restriction that drops the warm-up strengthens the estimate, and the ten-month
window is the end point of that sequence rather than an outlier in it.

\textbf{Placebo.} Splitting a single regime at a date on which no rule changed
should return a ratio of one. Within MR12 a split at October 2024 gives 0.981
[0.850, 1.139] and one at January 2025 gives 0.945 [0.816, 1.101]. Within MR15, on
the sample that begins in July 2022 and so excludes the warm-up quarters, splits at
January and March 2023 give 0.926 [0.749, 1.151] and 0.945 [0.789, 1.179]. Four
placebo comparisons, two in each regime, all covering one and none below 0.92. The
design does not manufacture declines.

\textbf{The choice of statistic.} Table 6 repeats the window comparison with the
measures the applied literature reports. Each is compared with its own length-only
benchmark, obtained by simulating the window's MR15 matchups through the calibrated
round process at fifteen and at twelve rounds.

\begin{table}[!ht]
\caption{The comparison under conventional measures, transition window.}
\label{tab:measures}
\centering
\footnotesize
\begin{tabular}{lccc}
\toprule
Measure & Observed ratio & Length-only benchmark & Below benchmark \\
\midrule
Discrimination $D$          & 0.667 & 0.938 & yes \\
Correlation $r$             & 0.766 & 0.940 & yes \\
$R^2$ (reliability)         & 0.587 & 0.884 & yes \\
Spearman's $\rho$           & 0.725 & 0.939 & yes \\
Kendall's $\tau_b$          & 0.726 & 0.946 & yes \\
Logit slope                 & 0.657 & 1.091 & yes \\
Area under the curve        & 0.961 & 1.013 & yes \\
\addlinespace
Noll--Scully ratio          & 0.956 & 0.940 & no \\
\bottomrule
\end{tabular}

\smallskip
\begin{minipage}{\textwidth}
\footnotesize
\emph{Notes.} All columns use the transition window (2,579 MR15 and 4,202 MR12
maps); every ratio is MR12 divided by MR15. The length-only benchmark simulates
the window's MR15 matchups through the calibrated beta-binomial round process at
fifteen and at twelve rounds with common random numbers, 120 replications, and
reports the ratio each measure would show if only the number of rounds had changed.
Cluster-bootstrap intervals over matches (300 replications) exclude one for six of
the seven association measures; the exception is the area under the curve. All estimates here are bivariate --- no
side indicator and no map fixed effects --- because these measures are
conventionally reported that way, which is why the $D$ ratio is 0.667 rather than
the controlled 0.648 of Table \ref{tab:calendar}. The area under the curve and the
logit slope use halves with a winner, dropping the 13.9 per cent of MR12 halves that
end six--six; a fifteen-round half cannot be drawn. The Noll--Scully ratio is the
exception and the informative one: it measures the dispersion of outcomes rather
than their association with ability, moves by exactly what fewer rounds imply, and
so would have registered nothing.
\end{minipage}
\end{table}

\begin{figure}[!ht]
\centering
\includegraphics[width=\textwidth]{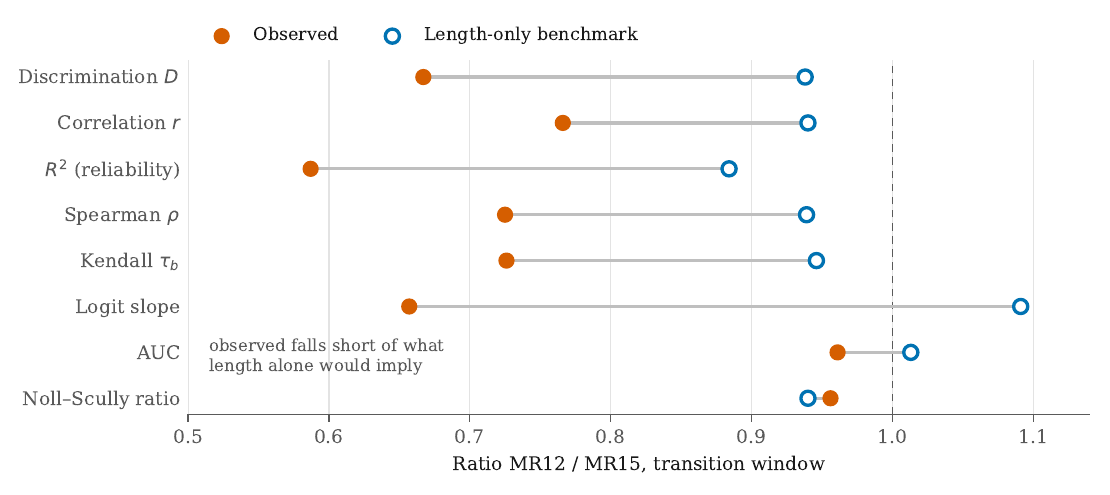}
\caption{The comparison under eight measures of association, transition window.
Filled circles are the observed MR12/MR15 ratio; open circles are the ratio the
same measure would show if only the number of rounds had changed. Every measure of
the association between ability and outcome falls short of its own length-only
benchmark. The Noll--Scully ratio, which measures the dispersion of outcomes rather
than their association with ability, does not.}
\label{fig:measures}
\end{figure}

Every one of them reproduces the result. The correlation falls to 0.766 of its MR15
value, R-squared to 0.587, Spearman's rho to 0.725, Kendall's tau-b to 0.726, the
logit slope to 0.657 and the area under the curve to 0.961, against 0.667 for $D$.
Each falls below its own length-only benchmark --- 0.940, 0.884, 0.939, 0.946,
1.091 and 1.013 respectively --- so the finding that the decline exceeds what fewer
rounds can deliver does not rest on the statistic we chose to feature.
Cluster-bootstrap intervals exclude one for six of the seven association measures: [0.566,
0.999] for the correlation, [0.321, 0.997] for R-squared, [0.526, 0.956] for
Spearman, [0.527, 0.960] for tau-b, [0.440, 0.919] for the logit slope and [0.498,
0.886] for $D$. The single exception is the area under the curve, whose interval of
[0.905, 1.008] covers one; it is also the measure most affected by a mechanical
asymmetry, since a twelve-round half can end six-six and a fifteen-round half
cannot, so 13.9 per cent of MR12 halves have no winner and what leaves the sample
is precisely the closest contests.

One conventional measure returns a precise null, and it should. The Noll-Scully
ratio of the observed to the idealised standard deviation of the favourite's
first-half round share falls from 1.403 to 1.341 in the window, a ratio of 0.956
against a length-only benchmark of 0.940. The standard competitive-balance
statistic moves by the amount fewer rounds imply and detects nothing beyond it.
This is the within-league form of the point \citet{doria2021} make about season
length, and it is why the balance ratio is not the outcome variable here.

\subsection{A Coincident Change: Side Balance}

The same estimates identify a structural change arriving with the same transition.
Under MR15, starting a half on the counter-terrorist side was worth 1.06 additional
rounds (standard error 0.10). Under MR12 the figure is $-0.01$ (standard error
0.06), indistinguishable from zero: the sides were balanced. Side advantage cancels
over a full map, since each team plays each side once, but it is a large source of
within-half variation, and the side indicator and its interaction with map fixed
effects absorb it throughout.

The result matters for interpretation rather than for the estimates. It shows
directly that the round-generating environment changed at the transition, not only
its length, and it is the identification problem of Section 4.4 made visible in the
estimates themselves. At least one accompanying change is large and measurable;
there may be others that are not.

\subsection{Limitations}

Four limitations bound the estimates. The first is identification: the round
format changed simultaneously with the engine, no measurement here separates them,
and the regime indicator is a joint indicator. We do not claim that shortening the
contest caused the observed fall in discrimination. We claim that the fall has the
sign the length mechanism predicts, that its magnitude in the window
comparisons exceeds what length alone can deliver under a calibrated round process,
and that at least one accompanying change, the equalisation of side balance, is
visible in the same estimates. The translation into win probabilities in Section
5.4 is an illustration computed from the fitted model, and inherits both that
limitation and the approximation error of the beta-binomial.

The second is that the window narrows calendar distance without eliminating it.
Inside the ten months the two formats are still largely sequential, and only three
of forty-three weeks contain both; the strictly contemporaneous comparison exists
but rests on 483 maps and its interval covers no change. Anything specific to the
autumn of 2023 and distinct from the format --- a patch, a break in the calendar, a
shift in which tier of event was running --- is therefore not differenced out. What
the design rules out is an era effect accumulating over years, not a shock lasting
a season.

The third is selection. The window comparison rests on a non-random set
of MR15 observations: leagues that chose to finish their seasons under the old
rules. Reweighting on observables and restricting to teams present in both regimes
do not overturn the contrast, but selection on unobservables cannot be ruled out.

The fourth is coverage and measurement. The ranking that defines the favourite
covers roughly the leading two hundred and fifty teams, so the results are
conditional on both teams being ranked and say nothing about the long tail of
professional play. The rating recursion makes no adjustment for roster changes,
which are frequent here and add classical measurement error to the ability
regressor; that error attenuates the slopes in both regimes.

\section{The Bias from Assuming Independent Trials}
\label{sec:bias}

The estimates above were built to test one rule change. The correction they rest on
is general, and this section states it in the form the tournament-design literature
can use.

Let a format aggregate $N$ trials whose pairwise correlation is $\rho$. The
effective number of independent trials is
\begin{equation}
N_{\text{eff}}(N,\rho) \;=\; \frac{N}{1 + (N-1)\rho},
\label{eq:neff}
\end{equation}
and discriminating power grows with its square root, so the efficacy of a format of
length $N_2$ relative to one of length $N_1$ is
$\sqrt{N_{\text{eff}}(N_2,\rho)/N_{\text{eff}}(N_1,\rho)}$ rather than
$\sqrt{N_2/N_1}$. Setting $\rho = 0$ recovers the independence benchmark that
simulation studies impose by construction.

The two expressions diverge sharply, and in a direction that matters for policy.
Under independence, cutting a format by twenty per cent always costs 10.6 per cent
of discriminating power, whatever the format's length: the ratio $\sqrt{0.8}$ does
not depend on $N$. Under dependence it depends on $N$ strongly, because the trials
removed from a long format were the ones carrying least new information. At the
intra-class correlation we estimate, $\hat\rho = 0.074$, a twenty per cent cut
costs 6.3 per cent at ten trials, 5.2 per cent at fifteen, 3.5 per cent at thirty
and 2.1 per cent at sixty. The independence assumption therefore overstates the
cost of shortening by a factor of 1.7 at ten trials and 5.1 at sixty.

\begin{table}[!ht]
\caption{What a twenty per cent cut in trials costs, under independence and under
the calibrated dependence.}
\label{tab:bias}
\centering
\footnotesize
\begin{tabular}{rrcccc}
\toprule
\multicolumn{2}{c}{Format length} & Effective trials & \multicolumn{2}{c}{Efficacy lost by the cut} & Overstatement \\
\cmidrule(lr){1-2}\cmidrule(lr){4-5}
from & to & after the cut & independent & calibrated & factor \\
\midrule
10 & \phantom{0}8 & 5.26 & 10.6\% & 6.3\% & 1.68 \\
15 & 12 & 6.60 & 10.6\% & 5.2\% & 2.02 \\
20 & 16 & 7.57 & 10.6\% & 4.5\% & 2.36 \\
30 & 24 & 8.86 & 10.6\% & 3.5\% & 3.04 \\
60 & 48 & 10.69 & 10.6\% & 2.1\% & 5.07 \\
\bottomrule
\end{tabular}

\smallskip
\begin{minipage}{\textwidth}
\footnotesize
\emph{Notes.} Each row cuts the number of trials by twenty per cent. ``Efficacy
lost'' is $1 - \sqrt{N_2/N_1}$ under independence and
$1 - \sqrt{N_{\text{eff}}(N_2,\rho)/N_{\text{eff}}(N_1,\rho)}$ under the
calibrated process, at $\hat\rho = 0.074$. The independence figure is 10.6 per
cent in every row because $\sqrt{0.8}$ does not depend on $N$; the calibrated
figure falls with length, because the trials a long format gives up were the ones
adding least. The final column is the ratio of the two, that is, the factor by
which an independence-based simulation overstates the cost of shortening.
\end{minipage}
\end{table}

Two implications follow for applied work.

First, published efficacy comparisons between formats of different length are
biased against the shorter format, and the bias grows with the length of the
formats compared. A simulation that treats games in a best-of-seven series, or
matches in a round-robin, as independent will report a larger advantage for the
longer design than the data support. The direction is not ambiguous: dependence
always moves the comparison towards no change, because $N_{\text{eff}}$ is concave
in $N$.

Second, the correction is cheap to apply. It needs one number, $\rho$, and that
number is recoverable from ordinary results data without observing the trials'
dependence structure directly: the ratio of the observed variance of a format's
score to its binomial counterpart is the design effect $\varphi = 1 + (N-1)\rho$
of survey sampling \citep{kish1965}, and $\sqrt{\varphi}$ is the Noll--Scully ratio
already familiar in this literature \citep{owen2010,doria2021}. Section 4.2 shows
the estimation in full; it requires only scores, and it must be computed
conditional on the strength gap, since unequal pairings otherwise enter as
spurious dependence.

A caution bounds both implications. Equation~\eqref{eq:neff} assumes exchangeable
trials, so a single $\rho$ describes every pair. That is a reasonable
approximation over the twelve- and fifteen-trial formats on which we calibrate it,
and it is what makes the ceiling $N_{\text{eff}} \to 1/\rho$ appear as $N$ grows;
but nothing in our data speaks to whether trials fifty apart are correlated like
trials one apart. Where dependence decays with distance --- plausible in long
seasons, less so within a single match --- the correction remains valid in
direction and should be recalibrated in magnitude.

\section{Concluding Remarks}

Efficacy comparisons between tournament formats of different length are biased when the trials a format aggregates are treated as independent, and this paper measures the bias. Because trials are serially correlated, a format carries $N/[1+(N-1)\rho]$ effective trials rather than $N$, so the cost of shortening is smaller than the independence benchmark implies and the gap widens with the length of the formats compared: a twenty per cent cut costs 10.6 per cent of discriminating power under independence, 5.2 per cent at the correlation we estimate, and 2.1 per cent in a sixty-trial format. The correction needs a single parameter, and that parameter is recoverable from ordinary results data as a design effect.

The ingredients come from an implemented change. The transition to Counter-Strike 2 and MR12 was accompanied by a fall in how sharply map outcomes track ability, and in the ten months around the switch that fall was larger than the calibrated model predicts. We do not attribute it to the round count: format and engine changed together, and Section 5.7 sets out what that leaves unidentified. Fifteen rounds carry the equivalent of 7.35 independent observations and twelve of the same process 6.60, so shortening on its own predicts a five per cent fall in the ability slope per unit of residual noise. We estimate ten per cent across 2022--2025, which is consistent with that benchmark, and thirty-five per cent in the window around the transition, which is not. The statistic behind these numbers is the ordinary discrimination parameter of measurement theory; what the paper adds is the benchmark it is measured against.

One implication follows for the evaluation of format changes. Shortening a contest cannot be assessed from duration, nor from the nominal number of trials removed: the informative object is how much ability-relevant information survives per unit of play, which depends on serial dependence in the scoring process as much as on the number of trials --- a rule change that halves the rounds does not halve the evidence.

The estimates measure a bundled transition, not the round count alone, because MR12 arrived with Counter-Strike 2. Two extensions would separate them. Formats within the discipline vary in length while leaving the round process untouched, and other esports have changed series length without changing the game; either provides the clean contrast this setting lacks. Whether the sharp contemporaneous decline persists as teams adapt to the new engine is a question the next three seasons of the same panel will answer.

\backmatter

\section*{Declarations}

\textbf{Funding.} This work/article is an output of a research project
HSE-BR-2025-82 implemented as a part of the Basic Research Program at HSE
University.

\textbf{Conflict of interest.} The author has no competing interests to declare
that are relevant to the content of this article.

\textbf{Ethics approval and consent to participate.} Not applicable. The study uses
public records of professional sporting results and involves no human or animal
subjects.

\textbf{Consent for publication.} Not applicable.

\textbf{Data availability.} The data underlying this article are derived from
public match records published by HLTV.org. A replication package containing the
collection pipeline, the parsers, the database schema and a compressed cache of
the retrieved pages is available from the author, so that every number in the
article can be reproduced without re-crawling the source. Appendix~\ref{secA1}
documents the crawl, the format-classification rules and the exclusions.

\textbf{Materials availability.} Not applicable.

\textbf{Code availability.} The estimation code forms part of the replication
package described above and is available from the author.

\textbf{Author contribution.} The author has accepted responsibility for the entire
content of this manuscript and approved its submission.

\textbf{Use of generative AI.} A generative AI assistant was used for drafting and
debugging analysis code, for language editing and for typesetting support. It was
not used to generate data, to select specifications or to interpret results. The
author verified every reported figure against the underlying data and takes full
responsibility for the content.

\begin{appendices}

\section{Data Collection and Sample Construction}\label{secA1}

This appendix documents how the panel was assembled, in enough detail for the
procedure to be repeated from the public source.

\subsection*{A.1 Source and crawl}

All records come from HLTV.org, the reference results archive for professional
Counter-Strike. Three page families are used. Match pages carry the date, the
event, the two teams, the series format, the per-map scores and the link to each
map's statistics page. Map statistics pages carry the half-by-half score in the
form \texttt{(a : b ; c : d)}, from which the first-half score and the starting
side of each team are recovered. Ranking pages carry the weekly world ranking and
its points, published each Monday, from which the pre-match ranking of both teams
is taken by the snapshot in force on the date of the match.

Pages were retrieved in chronological order of match identifier, one request at a
time with a fixed delay between requests, and cached to disk before parsing, so
that parsing is reproducible without re-crawling. The raw page cache is part of
the replication package. No page was retrieved more than once except where a
parse error required it.

\subsection*{A.2 Determining the round format}

The round format is the treatment, so it is read off each map rather than assigned
by date. Two signals are used, in that order of priority. First, the half-score
structure: a map whose two halves sum to fifteen rounds each is MR15, and one whose
halves sum to twelve is MR12. Second, where half scores are missing, the regulation
total: thirty rounds implies MR15 and twenty-four implies MR12. Maps for which
neither signal is available, or whose signals conflict, are flagged and dropped.

The source also carries a game-version field, but it is not an independent third
signal and we do not use it as one: in these records the version is itself inferred
from the round format, so treating it as corroboration would be circular. This is
the same point made in Section 3.2 --- the panel contains no marker of the engine
that is independent of the format --- and it is one reason the two changes cannot
be separated.

Reading the format from the map rather than the calendar matters because the
transition was staggered. Leagues already in progress on 1 September 2023 finished
their seasons under MR15 while new events started under MR12, and a small number of
events ran the two formats in parallel. It is precisely this staggering that
creates the ten-month window on which the paper's preferred comparison rests; a
date-based assignment would have destroyed it.

\subsection*{A.3 Exclusions}

Three exclusions define the estimation sample. Walkovers and forfeits are dropped
(1,446 records): a forfeit appears as a nominal 1--0 result on a map named
``Default'' with no round detail, and is not a contest. Maps for which the
pre-match world ranking of either team is unavailable are dropped (13,131 records),
because the favourite cannot be defined; these are overwhelmingly matches between
teams outside the published ranking. Maps whose round format cannot be determined
are dropped. What remains is the estimation sample of 58,294 maps from 26,958
matches, 1,094 teams and 1,934 events.

Specifications that use the half decomposition impose two further requirements:
the half-by-half score and the starting side must be recoverable, and both teams
must already have at least twenty-five maps in the rating recursion. This leaves
the 39,546 maps from 19,362 matches described in Table \ref{tab:desc}.

\subsection*{A.4 Variables}

The \emph{favourite} is the better-ranked team at the time of the match; ties in
rank are broken by ranking points and, failing that, the map is dropped. The
\emph{first-half margin} $M_i$ is the favourite's first-half rounds minus the
underdog's. The \emph{starting side} is the favourite's side in the first half,
read from the half-score structure. The \emph{rating gap} $S_i$ is the difference
in pre-match round-share Elo, favourite minus underdog, expressed per hundred
points; the recursion is described in Section 4.3 and its step size is examined in
Table \ref{tab:robust}.

\subsection*{A.5 Reproducibility}

The collection pipeline, the parsers, the database schema and the estimation code
are available in a replication package together with the compressed page cache, so
that every number in the paper can be reproduced without contacting the source
site. The package documents the crawl order, the request pacing and the
classification rules applied to each map.

\end{appendices}

\bibliography{refs}

\end{document}